\documentclass[prc,superscriptaddress]{revtex4}
\usepackage{amssymb,color,graphicx,bm,amsmath}

\definecolor{red}{rgb}{0.8,0,0}
\definecolor{RED}{rgb}{0.8,0,0}
\definecolor{violet}{rgb}{0.4,0,0.4}
\definecolor{green}{rgb}{0,0.5,0.0}
\definecolor{GREEN}{rgb}{0,0.5,0.0}
\definecolor{navy}{rgb}{0.0,0.0,0.6}
\definecolor{orange}{rgb}{0.8,0.2,0.0}
\definecolor{blue}{rgb}{0.3,0.0,0.8}

\newcommand{\lsim}{\raisebox{-.4ex}{$\stackrel{<}{\scriptstyle \sim}$}}

\begin{document}

\title{New mass limit for white dwarfs: super-Chandrasekhar type~Ia supernova
as a new standard candle}

\author{Upasana Das, Banibrata Mukhopadhyay\\
Department of Physics, Indian Institute of Science, 
Bangalore 560012, India\\ upasana@physics.iisc.ernet.in , bm@physics.iisc.ernet.in\\
}

\begin{abstract}
Type~Ia supernovae, sparked off by exploding white dwarfs of mass close to Chandrasekhar limit, play the key role to understand 
the expansion rate of universe. However, recent observations of several peculiar type~Ia supernovae argue for its
progenitor mass to be significantly super-Chandrasekhar. We show that strongly magnetized white dwarfs not only can violate the
Chandrasekhar mass limit significantly, but exhibit a different mass limit.
We establish from foundational level that the generic mass limit of white dwarfs is $2.58$ solar mass. This explains
the origin of over-luminous peculiar type~Ia supernovae. Our finding further argues for a possible second 
standard candle, which has many far reaching implications, including a
possible reconsideration of the expansion history of the universe.
\end{abstract}

\maketitle


\paragraph*{Introduction.$-$} Recently, some peculiar type~Ia supernovae: e.g. SN~2006gz,
SN~2007if, SN~2009dc, SN~2003fg, have been observed \cite{scalzo} with
exceptionally higher luminosities but lower kinetic energies. 
The kinetic energy is proportional to the difference between the obtained nuclear energy which
arises from the synthesis of elements in the explosion through fusion and the binding energy 
of the white dwarf. Most of the light-curves of above mentioned peculiar supernovae appear
over-luminous, slow-rising, which do not allow them to be calibrated 
as standard candles. This questions the use of all type~Ia supernovae in measuring distances of
far away regions and hence 
unraveling the expansion history of the universe. 
However, assuming the progenitor to be a highly super-Chandrasekhar mass white dwarf
reproduces the low kinetic energies
and thus velocities seen in the above supernovae \cite{nature}. This is
because a larger mass implies a larger binding energy of the star and hence a smaller
velocity for the same and/or higher luminosity (due to nuclear fusions) than that observed in
a standard type~Ia supernova. The progenitor masses required to explain the above
supernovae lie in the range $2.1-2.8M_\odot$, when $M_\odot$ being the mass of sun,
subject to the model chosen to estimate the nickel mass \cite{scalzo,nature,hicken,yam,silverman,
taub}.
Now naturally the following vital questions arise. Is there any fundamental basis behind 
the formation of a highly super-Chandrasekhar white dwarf? How to address the significant 
violation of Chandrasekhar mass limit?
What is the ultimate mass limit of a white dwarf?
Here we plan to address all the above issues by exploiting the effects of magnetic field
in compact objects. This will lead, as we will show, to a natural explanation of
the so called ``peculiar" type~Ia supernovae. This might eventually lead these supernovae to be 
considered as altogether new standard candles. This has many far reaching implications, including 
a possible reconsideration of the expansion history of the universe.

Before proceeding further, let us recall the physics of a white dwarf and its link
to the type~Ia supernova.
When a star exhausts its nuclear fuel, it converts to either of the three compact
objects: white dwarf, neutron star or black hole, depending on the initial 
mass of the evolving star. It is generally believed that the fate is a white dwarf
when the mass of the initial star in its main sequence is $M_{MS}\lsim 5M_\odot$. 
Such a main sequence star undergoing collapse 
leading to a small volume consists of a lot of electrons. Being in a
small volume many such electrons tend to occupy same energy states, hence making them
to degenerate electrons, as the
energy of a particle depends on its momentum which is determined by the total volume of 
the system. However, being a fermion an
electron obeys Pauli's exclusion principle which says that no two fermions can
occupy the same quantum state. Hence, once up to the Fermi level, which is the maximum 
allowed energy of a fermion, is filled by the electrons, there is no available space
for the remaining electrons in a small volume of a collapsing star, which expels the
electrons to move out leading to an outward pressure.
In a white dwarf, the inward gravitational force 
is balanced by the force due to outward pressure created by to degenerate electrons. 
Chandrasekhar in one of his celebrated
papers \cite{chandra35} showed that the mass of a white dwarf cannot be more than $1.44M_\odot$ 
which sets the famous Chandrasekhar mass limit of a white dwarf. 

If a white dwarf having mass close to the Chandrasekhar limit gains more mass (e.g. by accretion, 
when the mass is supplied by a companion star of the white dwarf),
then its mass exceeds the 
Chandrasekhar limit, which leads to a gravitational force stronger than the
outward force that arises due to the degenerate electrons. Hence, this leads 
to the contraction of the white dwarf and a subsequent increase of its temperature, 
which is favorable for the initiation of fusion reactions again. If the white dwarf
mostly consists of carbon and oxygen, namely
the carbon-oxygen white dwarf (which is commonly the case), then nuclear fusion of 
carbon (and oxygen) takes place. Subsequently, within a few seconds, a substantial fraction of 
the white dwarf matter undergoes a runaway reaction which releases huge energy $\sim 10^{51}$erg  
to unbind it in an explosion, namely type~Ia supernova explosion. 
This eventually leads to a complete gravitational
collapse of the star without leaving any remnant. 

As all the commonly observed type~Ia supernovae are produced
by (almost) the same mechanism, namely the mass of the progenitor white dwarf 
exceeding Chandrasekhar limit
and subsequent processes, the underlying variations of luminosity as functions of time,
namely light-curves, appear alike for all the explosions. All of these supernovae exhibit 
consistent peak luminosity, the relation between the peak luminosity and width of the 
light-curve \cite{phillips,gold}, 
due to the uniform mass of white dwarfs (Chandrasekhar limit) 
which finally explode, e.g., because of the accretion process. Very importantly, 
the apparent stability of this value helps the underlying supernovae to be used as 
standard candles in order
to measure the distances to their host galaxies. 
Since these supernovae are exceptionally bright, they can be observed across 
huge cosmic distances. 
The variation of their brightness with distance (or redshift) is an extremely 
important tool for measuring various 
cosmological parameters, which in turn shed light on the expansion history of the Universe.
Their enormous importance is self evident and was brought 
into the prime focus by the awarding of the Nobel Prize in Physics in 2011, for the discovery 
(made possible by the observation of distant type~Ia supernovae) that 
the universe is undergoing an accelerated expansion \cite{riess98,perl99}.

Now we move on to our goal of establishing a new mass limit for super-Chandrasekhar white dwarfs,
of whose formation there is no understanding from foundational level --- 
a caveat behind the hypothesis of super-Chandrasekhar progenitor for the peculiar 
type~Ia supernovae, raised by the earlier authors 
\cite{nature}. In fact they emphasized on the pursuit of theoretical studies in order to
assess the hypothesis.
Although, based on a numerical code for stellar binary evolution, the rotating white dwarfs are suggested  
to hold mass up to
$2.7M_\odot$ \cite{kato}, a foundational level calculation is missing and there is 
no estimate of mass limit of such a star either.

In this letter, we show that (highly) magnetized white dwarfs not only
can have mass $\sim 2.6M_\odot$, but also exhibit its ultimate limit of mass.
Hence, we propose a fundamentally new mass limit for white dwarfs,
which eventually helps in explaining light-curves of peculiar type~Ia supernovae.
This may further lead to establishing these supernovae as modified standard candles 
for distance measurement.

The motivation behind our approach lies in the discovery of several isolated magnetized white 
dwarfs through the Sloan Digital Sky Survey (SDSS)
with surface fields $10^5-10^9$G \cite{schmidt03}, \cite{vanlandingham05}.
Hence their expected central fields could be $2-3$ orders of magnitude higher.
Moreover, about $25\%$ of accreting white dwarfs, namely cataclysmic variables (CVs), are 
found to have high magnetic fields $10^7-10^8$G \cite{wick}.

\paragraph*{Equation of state.$-$}
As the starting choice is the magnetized white dwarf, we first recall
degenerate electrons under the influence of magnetic field which are known
to be Landau quantized \cite{lai}. Larger the magnetic field, smaller is the 
number of Landau levels occupied \cite{pathria} (see supplement \cite{sm}). 
Recent works \cite{dasmu,ijmpd,km} establish that Landau quantization 
due to strong magnetic field modifies the equation of state (EoS) of the electron degenerate gas, 
which results in a significant modification of the the mass-radius relation of the 
underlying white dwarf. Interestingly, these white dwarfs 
are found to have super-Chandrasekhar masses.
The main aim of this letter
is to obtain the maximum possible mass of such a white dwarf (which is magnetized), 
and therefore a (new) mass limit. Hence we look for the regime of
high density of electron degenerate gas and the corresponding EoS, 
which further corresponds to the high Fermi energy ($E_F$) of the system. This is
because high density corresponds to high momentum, which implies high 
energy (see supplement \cite{sm}). Note that the maximum Fermi energy ($E_{Fmax}$) corresponds to
the maximum central density of the star. Consequently, conservation of magnetic flux 
(technically speaking flux freezing theorem which is generally applicable for a compact star)
argues for the maximum possible field of the system, which implies that only the
ground Landau level will be occupied by the electrons. For the expressions of
density, pressure and the EoS for such a highly magnetized system, see supplement \cite{sm}.
Hence, in the limit of $E_F>>m_ec^2$, when $m_e$ is the mass of the electrons and $c$ the 
speed of light, for
a given magnetic field exhibiting the system of one Landau level, 
the EoS is 
\begin{eqnarray}
P=K_m~\rho^2,
\label{eos}
\end{eqnarray}
when $P$ and $\rho$ are respectively the pressure and density of the gas and 
the constant $K_m$ is given by 
\begin{eqnarray}
K_m=\frac{m_ec^2\pi^2\lambda_e^3}{(\mu_em_H)^2B_D},
\label{kk}
\end{eqnarray}
where $\lambda_e=\hbar/m_ec$, the Compton wavelength of electron, $\hbar$ the Planck's constant $h$ divided by $2\pi$,
$\mu_e$ the mean molecular weight per electron, $m_{H}$ the mass of hydrogen atom and 
$B_D$ the magnetic field in the units of $4.414\times 10^{13}$G.

\paragraph*{Mass limit of white dwarfs.$-$}
Now following the Lane-Emden
formalism \cite{arc}, we obtain the 
mass of the magnetized white dwarf
\begin{eqnarray}
M=4\pi^2\rho_c\left(\frac{K_m}{2\pi G}\right)^{3/2},
\label{mas2}
\end{eqnarray}
and the corresponding radius
\begin{eqnarray}
R=\sqrt{\frac{\pi K_m}{2G}},
\label{rad2}
\end{eqnarray}
when $\rho_c$ is the central density of the white dwarf supplied as a boundary condition
in addition to the condition that $d\rho/dr=0$ at $r=0$, when $r$ is the radial distance from 
the center of the star such that at the surface $r=R$, and $G$ is the Newton's
gravitation constant. See supplementary information \cite{sm} for detailed calculations.

Now the expression of $\rho_c$ for a one Landau level system in the limit 
$E_F=E_{Fmax}>>m_ec^2$ is given by (see supplement \cite{sm})
\begin{eqnarray}
\rho_c=\frac{\mu_em_H}{\sqrt{2}\pi^2\lambda_e^3}B_D^{3/2} .
\label{rhoc}
\end{eqnarray}
Substituting $\rho_c$ from equation (\ref{rhoc}) in equation (\ref{mas2}), we obtain 
the mass of the white dwarf, independent of $\rho_c$ and $B_D$, given by
\begin{eqnarray}
M=\left(\frac{hc}{2G}\right)^{3/2}\frac{1}{(\mu_em_H)^2}\approx\frac{10.312}{\mu_e^2}M_\odot,
\label{mas3}
\end{eqnarray}
and from equations (\ref{kk}), (\ref{rad2}) and (\ref{rhoc}) we obtain the radius
\begin{eqnarray}
R=\left(\frac{\pi^{2/3}~hc}{2^{7/3}~(\mu_em_H)^{4/3}~G}\right)^{1/2}\rho_c^{-1/3}\rightarrow 0\,\,\,
{\rm as}\,\,\,\rho_c\rightarrow\infty,
\label{rad3}
\end{eqnarray}
which set the new limits for mass and radius.
Note that the Chandrasekhar limit also corresponds to $\rho\rightarrow\infty$ and $R\rightarrow 0$.
For $\mu_e=2$ which is the case of a carbon-oxygen white dwarf
\begin{eqnarray}
M\approx2.58M_\odot.
\label{mas4}
\end{eqnarray}
Interestingly, while high magnetic field introduces anisotropic effects into the star tending
it to be an oblate spheroid, this does not affect the limiting mass as the corresponding 
radius tends to zero. However, for lighter white dwarfs with finite radii, super-Chandrasekhar
mass would have been achieved at a lower field, if the star is appropriately set to be a spheroid rather
than a sphere, as justified earlier \cite{dasmu,ijmpd}.

\paragraph*{Scaling behaviors of mass and radius of the white dwarf with its central density.$-$}
Now we provide general scaling laws for the mass and radius of the white dwarf
describing their variations with its central density (as is known for non-magnetized 
white dwarfs proposed by Chandrasekhar) and magnetic field strength.
By Lane-Emden formalism
\begin{equation}
M \propto K^{3/2} \rho_c^{\frac{3-n}{2n}},
\label{msc}
\end{equation}
where $K$ depends on $B_D$ and $\rho_c$ for a magnetized white dwarf (see supplement \cite{sm}). 
For the extremely high density regime of the white dwarf, 
the EoS reduces to a polytropic form $P=K\rho^{\Gamma}$ 
(can be verified from supplement \cite{sm}), when $\Gamma=1+1/n$, is the polytropic index. 
In this regime, $\Gamma=2$, consequently $n=1$  and $K=K_m\propto B_D^{-1}\propto\rho_c^{-2/3}$, 
as shown by equations (\ref{eos}), (\ref{kk}), (\ref{rhoc}).
This finally renders $M$ in equation (\ref{msc}) to be
independent of $\rho_c$, as already shown by equation (\ref{mas3}). Similarly, the scaling of
radius is obtained as
\begin{equation}
R \propto K^{1/2} \rho_c^{\frac{1-n}{2n}},
\label{rsc}
\end{equation}
which reveals $R\propto\rho_c^{-1/3}$ for $n=1$, as already shown by equation (\ref{rad3}). 

\paragraph*{Justification of high magnetic field in white dwarfs.$-$}
So far we have employed Landau quantization in search of a new mass limit giving rise to
super-Chandrasekhar white dwarfs. However, the effect of Landau quantization becomes significant
only at a high field $\sim B_{D}\times 10^{13}{\rm G}= B_{cr}$. How can we justify such a high field in a white dwarf? 

Let us consider the commonly observed phenomenon of a magnetized white dwarf accreting mass from its companion. 
Now the surface field of an accreting white dwarf, as observed, could be $\sim 10^9{\rm G}<<B_{cr}$ \cite{schmidt03}. 
Its central field however can be several orders of magnitude higher 
$\sim 10^{12}$ G, which is also less than $B_{cr}$.
Naturally, such a magnetized CV, commonly known as a polar, still lies on the mass radius relation obtained 
by Chandrasekhar. However, in contrast with Chandrasekhar's work (which did not include magnetic field in 
the calculations), we will see that a non-zero initial field in the white dwarf, however ineffective for rendering 
Landau quantization effects, will 
prove to be crucial in supporting the additional mass accumulated due to accretion. As the magnetized white 
dwarf accretes mass, its total mass increases which in turn increases the gravitational power and hence the 
white dwarf contracts in size due to the increased gravitational pull. However, the total magnetic 
flux is conserved in such a process and, hence, as a result of the above decrease in size of the star, 
the central (as well as surface) magnetic field also increases. 
Here we are interested in the evolution of the central field, since it is 
this field which is primarily responsible for rendering 
super-Chandrasekhar mass to the white dwarf, as justified in \cite{dasmu}.
Since accretion is a 
continuous process, the deposition of matter on the surface of the white dwarf, followed by its contraction and subsequent 
increase of magnetic field, continues in a cycle. In such a process, eventually the central magnetic field 
could exceed $B_{cr}$. As a result,
the EoS of the electron degenerate matter gets modified as shown in Figure 1. Hence,
the inward gravitational force is balanced by the outward force due to this modified pressure and a 
quasi-equilibrium state is attained. In this way, a very 
high magnetic field is generated, which in turn prevents
the white dwarf from collapsing, thus making it more massive. Subsequently, with the continuation of accretion
the white dwarf approaches the new mass limit $\sim 2.58 M_\odot$, as obtained above, which sparks off 
a violent thermonuclear reaction with further accretion, thus exploding it and giving rise to a 
super-Chandrasekhar type~Ia supernova. 
The evolution of the mass-radius relation of a polar into that of a super-Chandrasekhar white dwarf
of the maximum possible mass is shown in Figure 2, 
along with a few typical mass-radius relations for different magnetic field strengths describing possible stars
in intermediate equilibrium states. 
The ultimate white dwarf, corresponding 
to the maximum mass $\sim 2.58M_{\odot}$, lies on the mass-radius 
relation for a one Landau level system, but the intermediate white dwarfs having weaker 
magnetic fields correspond to multilevel systems. 
The one Landau level system corresponds to the central magnetic field $8.8\times 10^{17}$~G.
The intermediate systems of 200-level and 50124-level correspond to 
central magnetic fields $4.4\times 10^{15}$~G and 
$1.7\times 10^{13}$~G 
respectively (see supplement \cite{sm} for relevant formula), when $E_{Fmax}=200m_ec^2$.
This value of $E_{Fmax}$ is found to produce the theoretical mass limit with good numerical accuracy.
Note that one can in principle go upto higher $E_{Fmax}$, which will lead to a further 
decrease in the radius of the white dwarf keeping the mass practically same. 
In order to construct this evolutionary track, we consider 
the values of $\rho_c$ corresponding to the density at the ground-to-first Landau level transition of the 
respective EoSs, since only this choice leads to the maximum possible mass.
We emphasize here that the range of masses for the super-Chandrasekhar progenitors obtained from observations is not 
very strict. Hence, if the new mass limit obtained by us is taken into account, one could possibly do away with 
the mass distribution altogether. However, if the distribution is indeed real, then it could be attributed to the 
difference in accretion rates found in different CVs.

We thank A.R. Rao of TIFR for insightful suggestions. B.M. acknowledges partial 
support through research Grant No. ISRO/RES/2/367/10-11. U.D. thanks CSIR, India for financial support.

\begin{figure*}
\begin{center}
\includegraphics[angle=0,width=7.5cm]{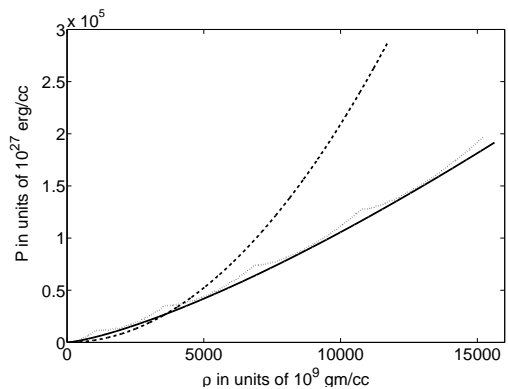}
\caption{ 
Equations of state --- the solid line
represents Chandrasekhar's equation of state (corresponding to
zero magnetic field and hence infinitely many Landau levels), the dotted and dashed lines represent the 
5-level (corresponding to very
strong magnetic field) and 1-level (corresponding to ultimate equation of state for
extreme magnetic field) systems of Landau quantization respectively. $E_{Fmax}=200m_ec^2$. 
  }
\label{eos1}
\end{center}
\end{figure*}

\begin{figure*}
\begin{center}
\includegraphics[angle=0,width=7.5cm]{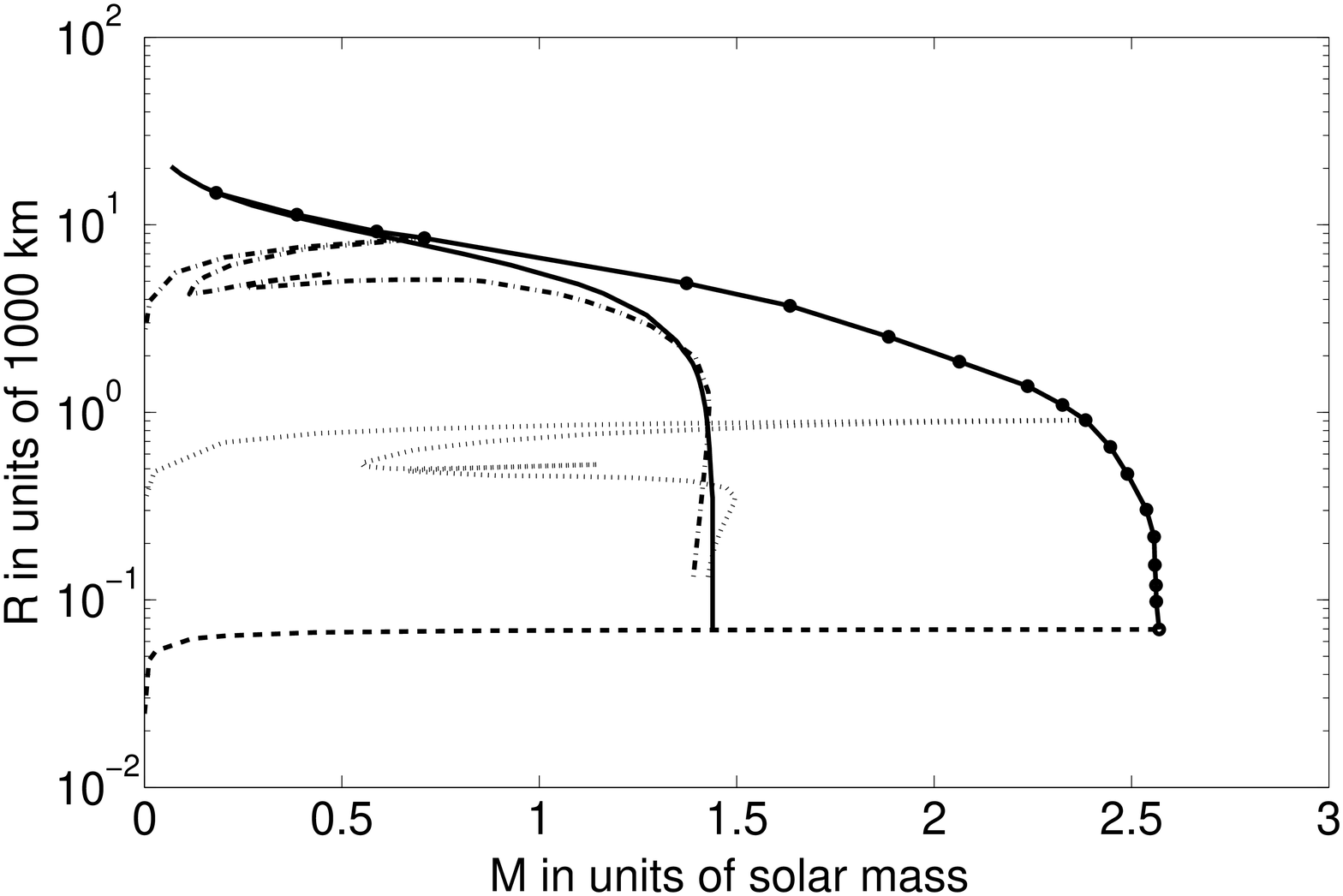}
\caption{ 
Mass-Radius relations --- the pure solid line
represents Chandrasekhar's result and the one marked with filled circles represents the evolutionary 
track of the white dwarf with the increase of magnetic field. The dot-dashed, dotted and dashed 
lines represent the white dwarfs with 50124-level, 200-level and 1-level
systems of Landau quantization respectively (corresponding to increasing magnetic fields). $E_{Fmax}=200m_ec^2$.
  }
\label{eos1}
\end{center}
\end{figure*}

\newpage
\numberwithin{equation}{section}
\section*{\large Supplementary information for the letter}

\section{Basic equations for one Landau level system of degenerate electrons}

The Fermi energy level ($E_F$) of a Landau quantized system is given by
\begin{eqnarray}
E_{F}^{2} = p_{F}(\nu)^{2}c^{2} + m_{e}^{2}c^{4}\left(1 + 2\nu B_{D}\right),
\end{eqnarray}
when $p_F(\nu)$ is the Fermi momentum of $\nu$th Landau level ($\nu=0,1,2.....$), $m_e$ the 
mass of electrons, $c$ the speed of light, $B_D$
the magnetic field in units of $4.414\times 10^{13}$G. 
From the condition that $p_{F}(\nu)^{2} \geq 0$, the maximum number of occupied Landau
levels is given by \cite{lai}
\begin{equation}
\nu_m= \frac{\left(\frac{E_{Fmax}}{m_ec^2}\right)^{2} - 1}{2B_{D}}. 
\label{be}
\end{equation}
For one Landau level system, when only ground Landau level ($\nu=0$) is occupied, $\nu_m=1$.
Similarly, for two level system, when ground ($\nu=0$) and first ($\nu=1$) levels are occupied, 
$\nu_m=2$, and so on. Hence, for one Landau level system, 
density ($\rho$) and pressure ($P$) 
of the electron degenerate gas \cite{km} are given by
\begin{eqnarray}
\rho=\frac{\mu_em_H B_D}{2\pi^2\lambda_e^3}\sqrt{\left(\frac{E_F}{m_ec^2}\right)^2-1}\longrightarrow 
\frac{\mu_em_H }{2\pi^2m_e c^2\lambda_e^3}B_DE_F\,\,\,{\rm for}\,\,\,E_F>>m_ec^2,
\label{rho}
\end{eqnarray}
\begin{eqnarray}
P=\frac{B_Dm_ec^2}{4\pi^2\lambda_e^3}\left(\frac{E_F\sqrt{E_F^2-m_e^2c^4}}{(m_ec^2)^2}
-\ln\left\{\frac{E_F+\sqrt{E_F^2-m_e^2c^4}}{m_ec^2}\right\}\right),
\label{pres}
\end{eqnarray}
when $\mu_e$ is the mean molecular weight, $m_H$ the mass of proton, 
$\lambda_e$ the Compton wavelength of electron.
Now eliminating $E_F$ from equations (\ref{rho}) and (\ref{pres}) for an arbitrary $E_F$, we arrive at the equation 
of state (EoS) for a one level system given by
\begin{eqnarray}
P=\frac{m_ec^2}{2Q\mu_em_H}\left(\rho\sqrt{Q^2+\rho^2}-Q^2\ln\left\{\frac{\rho+\sqrt{Q^2+\rho^2}}
{Q}\right\}\right),
\label{eos}
\end{eqnarray}
when $Q=\mu_e m_H B_D/2\pi^2\lambda_e^3$.

Following previous work \cite{dasmu}, we now approximate the above EoS and EoSs for any other $\nu_m$ 
by a polytropic relation 
$P=K\rho^\Gamma$ such that the polytropic
index $\Gamma=1+1/n$ is piecewise constant in different density ranges  and $K$ being a dimensional 
constant. This will prove to be useful in order to understand 
scaling behaviors of mass and radius of the white dwarf with its central density.

Now in terms of $Q$, equation (\ref{rho}) can be rewritten as
\begin{eqnarray}
\rho=Q \frac{E_F}{m_e c^2} .
\label{Qrho}
\end{eqnarray}
In the limit $E_F>>m_ec^2$, we note that, $\rho >> Q$. 
Thus in this limit, equation (\ref{eos}) reduces to
\begin{equation}
P=\frac{m_ec^2}{2Q\mu_em_H}\left(\rho^2-Q^2\ln\left\{\frac{2\rho}{Q}\right\}\right) . 
\end{equation}
The logarithmic term is much smaller than the first term in the above equation and hence by 
neglecting it we obtain 
\begin{equation}
P=\frac{m_ec^2}{2Q\mu_em_H}\rho^2,
\end{equation}
which corresponds to the polytropic EoS with $\Gamma=2$.

\section{Lane-Emden equation and expressions for mass and radius}

The underlying white dwarf following above EoS obeys the magnetostatic equilibrium 
condition 
\begin{eqnarray}
\frac{1}{\rho}\frac{d}{dr}\left(P+\frac{B^2}{8\pi}\right)=F_g+\frac{\vec{B}\cdot\nabla\vec{B}}{4\pi\rho},
\label{magstat}
\end{eqnarray}
when $r$ is the radial distance from the center of white dwarf, $\vec{B}$ the magnetic field 
in G, $B^2=\vec{B}\cdot\vec{B}$, $F_g$ the gravitational force. This equation is supplemented
by the estimate of mass ($M$) within any $r$ given by
\begin{eqnarray}
\frac{dM}{dr}=4\pi r^2\rho
\label{mas}
\end{eqnarray}
approximating the star to be spherical.
The white dwarf in the present context can be considered in the Newtonian 
framework and the magnetic field therein is nearly constant in the 
regime of smaller radii (as justified previously \cite{dasmu}). Moreover,
at a very large density, as in the limiting case to be considered here, the star
becomes so compact as if the magnetic field remains constant throughout.
Hence, taking above facts into consideration and combining equations (\ref{magstat}) and (\ref{mas}),
the equilibrium condition may be read
at any $r$ as
\begin{equation}
\frac{1}{r^{2}}\frac{d}{dr}\left(\frac{r^{2}}{\rho}\frac{dP}{dr} \right) = -4\pi G \rho,
\label{diff}
\end{equation}
where $G$ is the Newton's gravitation constant.
We now briefly recall the Lane-Emden formalism (see, e.g., \cite{arc}). Let us define
\begin{equation}
\rho = \rho_{c}\theta^{n},
\label{theta}
\end{equation}
where $\rho_{c}$ is the central density of the white dwarf and $\theta$ is a dimensionless variable
and
\begin{equation}
r = a\xi,
\label{xi}
\end{equation}
where $\xi$ is another dimensionless variable and constant $a$ carries the dimension of 
length defined as
\begin{equation}
a = \left [\frac{(n+1)K\rho_{c}^{\frac{1-n}{n}}}{4\pi G} \right ]^{1/2}.
\label{a}
\end{equation}
Thus using equations (\ref{theta}), (\ref{xi}) and (\ref{a}) along with the polytropic
form of EoS, equation (\ref{diff}) reduces to 
\begin{equation}
\frac{1}{\xi^{2}}\frac{d}{d\xi}\left(\xi^{2} \frac{d\theta}{d\xi} \right) = - \theta^{n},
\label{lane}
\end{equation}
which is the famous Lane-Emden equation.
Equation (\ref{lane}) can be solved for a given $n$, along with the boundary conditions
\begin{equation}
\theta(\xi = 0) = 1
\label{bc1}
\end{equation}
and
\begin{equation}
\left (\frac{d\theta}{d\xi} \right)_{\xi=0} = 0.
\label{bc2}
\end{equation}
Note that for $n < 5$, $\theta$ becomes zero for a finite value of $\xi$, say $\xi_{1}$, 
which basically corresponds to the surface of the white dwarf such that its radius 
\begin{equation}
R = a\xi_{1}.
\label{R}
\end{equation}
Also by combining equations (\ref{mas}), (\ref{theta}), (\ref{xi}) and (\ref{lane})  we obtain the 
mass of the white dwarf
\begin{equation}
M = 4\pi a^{3} \rho_{c}\int \limits_{0}^{\xi_{1}} \xi^{2}\theta^{n}\, d\xi.
\label{M}
\end{equation}


\end{document}